\documentclass[twocolumn,showpacs,prb,amsmath,amssymb]{revtex4}

\usepackage{graphicx}
\usepackage{bm}
\usepackage{mathbbol}

\DeclareMathAlphabet{\mathpzc}{OT1}{pzc}{m}{it}
\begin{document}

\title{Nuclear spins as quantum memory in semiconductor nanostructures}
\author{W. M. \surname{Witzel}} 
\author{S. \surname{Das Sarma}}
\affiliation{Condensed Matter Theory Center,
  Department of Physics, University of Maryland, College Park, Maryland
  20742-4111, USA} 
\date{\today}
\begin{abstract}
We theoretically consider the possibility of using solid state nuclear
spins in a semiconductor nanostructure environment as long-lived,
high-fidelity quantum memory.  In particular, we calculate,
in the limit of a strong applied magnetic field,
the fidelity of P donor nuclear spins as a function of time
in random bath environments of Si and GaAs, 
and the lifetime of excited intrinsic
spins in polarized Si and GaAs environments.
In the former situation, the nuclear spin dephases due to spectral
diffusion induced by the dipolar interaction among nuclei in the 
bath; in the interest of the high fidelity requirements
necessary for fault-tolerant quantum computing, we consider the 
{\it initial} quantum memory decay caused by a non-Markovian bath.
We calculate this nuclear spin memory time in the context of 
Hahn and Carr-Purcell-Meiboom-Gill refocused spin echoes using
a formally exact cluster expansion
technique which has previously been successful in dealing with {\it
  electron} spin dephasing in a solid state nuclear spin bath.  
With decoherence dominated by transverse dephasing ($T_2$),
we find it feasible to maintain high fidelity 
(losses of less than $10^{-6}$) quantum
memory on nuclear spins for times of the order
of $100~\mbox{$\mu$s}$ (GaAs:P) and $1$-$2~\mbox{ms}$
(natural Si:P) using CPMG pulse sequences of just a few
($\sim 2-4$) applied pulses.  
We also consider the complementary situation of a central
flipped intrinsic nuclear spin in a bath of completely polarized nuclear spins
where decoherence is caused by the direct flip-flop of the central
spin with spins in the bath.  Exact numerical calculations that include
a sufficiently large neighborhood of surrounding nuclei show lifetimes 
on the order of $1$-$5~\mbox{ms}$ for both GaAs and natural Si.  
Our calculated nuclear spin coherence times may have significance for
solid state quantum computer architectures using localized electron
spins in semiconductors where nuclear spins have been proposed for
quantum memory storage.
\end{abstract}


\maketitle

\section{Introduction}
\label{introduction}
The motivation for developing a solid state quantum computer
architecture using localized spins as qubits arises primarily from the
long presumed quantum coherence times for spins even in the strongly
interacting solid state environment.  In this respect, nuclear spins
are ideal since both spin relaxation (i.e., $T_1$) and spin coherence
(i.e., $T_2$) times are very long for nuclear spins, as compared with
electron spins, due to their weak coupling to the environment.  
The application of a strong
magnetic field further enhances nuclear coherence by suppressing, at
least, the leading order relaxation and decoherence processes 
caused by direct hyperfine coupling between nuclear spins and any
surrounding electron spins due to the large mismatch between electron
and nuclear spin Zeeman energies.  Throughout this paper we assume a
strong applied magnetic field that defines the
$z$-direction of our interacting spin system, and we also neglect,
somewhat uncritically, all effects of any direct hyperfine coupling
between electron spins and nuclear spins assuming our system to be
entirely a nuclear spin system.
The existence of localized electron spins in the environment will
further suppress the nuclear spin coherence, and therefore our
theoretical values for nuclear spin quantum memory lifetimes should
be taken as upper bounds.

Solid state nuclear spins as possible qubits and/or quantum memory 
have earlier been discussed in the 
literature\cite{KaneNature98, LaddPRB05, ChildressPRL06,
  vanLoockPRL06,Taylor03}
in various contexts although we are not aware of any concrete
calculations of spectral diffusion-induced nuclear spin decoherence 
using the cluster expansion technique
(or any other quantitative techniques for that matter) 
along the line we carry out in this work.  
In particular, the vast NMR literature\cite{generalNMRlit}
(and the more recent literature on the prospective use of nuclear
spins in the quantum information processing\cite{NielsonChuang}) 
typically (and uncritically)  assume exponential Markovian decay of
nuclear spin quantum memory whereas we theoretically treat the 
non-Markovian nonexponential decay which dominates the initial
decoherence of interest to quantum memory applications that
require high fidelity.  We emphasize
in this context that we use the term $T_2$ to signify the time scale
of the initial decay behavior 
(defined as the time in which the extrapolation of this initial
decay behavior reaches $1/e$).
This is in sharp contrast to the ordinary
definition of 
$T_2$ where an exponential decay is implicitly assumed.


The theoretical work presented in this paper deals entirely with
the quantum 
memory of nuclear spins in solid state spin quantum
computer architectures.
  In particular, we directly calculate solid
state nuclear spin decoherence in the nuclear spin bath environment,
i.e. for a Hamiltonian which contains {\it only} interaction between
nuclear spins.  
Our work is of direct relevance to semiconductor-based
quantum computer architectures where the idea of using nuclear spins
to coherently store quantum spin memory has been seriously proposed in
the literature.\cite{KaneNature98, LaddPRB05, ChildressPRL06,
  vanLoockPRL06, Taylor03}
As such, although we develop a rather general
theoretical formalism for nuclear spin decoherence, we concentrate on
GaAs quantum dots and phosphorus doped silicon, Si:P, for applying our
theory since much of the solid state quantum computer work in the
literature has focused on these two architectures.\cite{GaAsSiP}

The theoretical
question we address is simple and well-defined: How long is the 
(high-fidelity) solid state quantum memory time for nuclear spins in 
semiconductor structures?  We answer this question in some detail for 
our specific model and systems, and discuss, using our formalism, how one can
enhance the nuclear spin quantum memory (beyond the simple Hahn spin
echo situation) by using composite pulse sequences to refocus nuclear
spins.  Our goal is thus a qualitative and quantitative understanding
of nuclear spin quantum memory in GaAs quantum dots and Si:P doped
structures taking into account the complex interacting environment of
all the other nuclear spins invariably present in the solid state
environment.

In considering nuclear spins as solid state quantum memory, we discuss
two inequivalent (and perhaps complementary) situations.  The usual
situation is a bath of unpolarized nuclear spins surrounding the
``central'' donor nuclear spin which serves as our quantum memory.  In this
situation, spin relaxation processes ($T_1$) are suppressed by applying
a strong magnetic field assuming there is a significant mismatch
between magnetic moments of the donor and intrinsic nuclei. 
Instead, the central spin memory dephases ($T_2$)
primarily through the spectral diffusion mechanism
that is induced by the surrounding nuclear spin bath.
In the spectral diffusion process for the high-fidelity (initial
decay) regime that is our focus, the central spin quantum memory is effectively
subjected to a non-Markovian temporally fluctuating magnetic field due
to the nuclear spin flip-flops in the surrounding bath that result from
dipolar interactions among the bath nuclei.  The
spectral diffusion, examined in Sec.~\ref{SD_nuc_mem},
sets an ultimate limit to the coherence time for 
nuclear quantum memory storage in an unpolarized bath.  
There are, however, well-established procedures, going back
fifty years,\cite{Meiboom, Waugh68}
which can enhance (or more accurately restore) the
quantum coherence over longer times.  These coherence-enhancement
protocols involve intricate refocusing of the nuclear spin dynamics
through well-designed composite pulse sequences.  We discuss a
particular kind of coherence enhancement protocol, namely the
Carr-Purcell-Meiboom-Gill (CPMG) composite pulse sequence,
\cite{Meiboom}
in the context of spectral diffusion induced nuclear spin memory decoherence
in Sec.~\ref{CPMG} of this paper.  In Sec.~\ref{nuc_transport}, we
consider the complementary situation with quantum memory stored in a
``central'' intrinsic spin in a bath that is
{\it completely} spin-polarized (perhaps through a dynamic nuclear
polarization technique).
This somewhat
artificial situation (because complete nuclear spin polarization of
the surrounding bath is highly improbable) is instructive because
spectral diffusion is, by design, completely suppressed in a fully
spin polarized nuclear spin bath (having no spin degrees of freedom
available to cause fluctuations); relaxation, however, is now possible via
direct dipolar flip-flops between the central nucleus and like nuclei
(those with the same magnetic moment as the central nucleus and are
thus still allowed to flip-flop in the limit of a strong applied
magnetic field) in the bath.
We solve this problem with exact quantum simulation (greatly
simplified because of the complete polarization of the bath) and discuss the
results in Sec.~\ref{nuc_transport}.
We conclude in Sec.~\ref{conclusion} with a discussion of the
implications of our results for solid state quantum information
processing and of the various limitations of our theory.

\section{Spectral Diffusion of a Donor Nucleus}
\label{SD_nuc_mem}

Nuclear storage of quantum information in a solid state environment is
most naturally placed on donor nuclei that are easily distinguishable
from the surrounding intrinsic nuclei.  
It is imperative that the memory is stored in a nucleus
which is distinct from the surrounding nuclei in some manner so that
the stored information can be recovered.
Several quantum computing architecture 
proposals\cite{KaneNature98,LaddPRB05,ChildressPRL06,vanLoockPRL06}
exploit the long-term quantum information storage capabilities
which donor nuclei spins can possess.
In this section, we present theoretical calculations of the $T_2$ dephasing of
donor nuclear spins in two solid state environments of interest for
quantum computing: Si:P and GaAs:P.  Specifically, we present
coherence versus time information in the context of simple Hahn echo
refocusing; in the next section, we address multiple pulse
Carr-Purcell-Meiboom-Gill (CPMG) refocusing.

Fault-tolerant quantum computation demands high fidelity; quantum
error correction typically requires $10^4$ to $10^6$ coherent 
operations~\cite{QEC_threshold} and therefore quantum memory loss
should be kept below $10^{-4}$ to $10^{-6}$.
The decoherence of nuclear spins has been well-studied in nuclear
magnetic resonance (NMR) going back 50 
years\cite{generalNMRlit, NMRlit} and more recently.\cite{LaddPRB05, recentNMR}
However, these
studies were not motivated by the need for such high fidelity.
In NMR, it is typical and appropriate to assume Markovian-type decoherence,
with an exponential decay form.  However, we find
that the initial $10^{-6} - 10^{-4}$ part of the decay of nuclear spin
coherence is non-Markovian (nonexponential).  For our purposes, then,
$T_2$ must take on a slightly different meaning than that of NMR.  We find
that the initial spectral diffusion decay will be of the form
$\exp{\left[-\left(t / T_2\right)^n\right]}$, defining $T_2$ as a
characteristic time scale for the initial decay (it is defined as
the extrapolated $1/e$ time of the decay 
although the decay will not generally match this extrapolation 
because it will eventually cross over into the Markovian regime
with an exponential form).  For considerations of high-fidelity
quantum memory storage, the $n$ exponent in the exponential holds as
much relevance as the $T_2$ time itself.

We have previously reported\cite{witzelHahn} on
a cluster expansion technique that is
often applicable to spectral diffusion problems, giving formally exact
solutions for short times.  In some systems, formally exact solutions
are obtained out to times that are longer than the decay time of the
system; in such cases, the cluster expansion method solves the entire
problem for practical purposes because the tail of the decay is of no 
consequence (assuming a monotonic decrease in coherence with time).  Other
systems, including the donor nucleus spin decay currently being
investigated, yield convergent cluster expansion results for only the
initial part of the decay.  For this reason, and motivated by the
strict high-fidelity requirements of quantum error correction,
we focus on the initial
decay and our plots show memory loss as a function of time rather than
exhibiting full, formally exact, decay curves.

In our model, we take the limit of a strong applied magnetic field
which completely suppresses any interactions that do not conserve
Zeeman energy.  For simplicity, we also disregard
interactions with any electrons; our Hamiltonian is solely composed of
interactions between nuclei.  We consider only the spectral diffusion
decoherence mechanism due to the dipolar flip-flops in the
nuclear bath surrounding the central nuclear spin.
The bath Hamiltonian consists of dipolar flip-flop interactions
between like nuclei (as well as the diagonal $I_{nz} I_{mz}$ interaction):
\begin{eqnarray}
\label{H_B}
{\cal H}_B &\approx& \sum_{n \ne m} b_{nm} \left[
\left\{
\begin{array}{cl}
I_{n+} I_{m-}& \mbox{if}~\gamma_n = \gamma_m \\
0& \mbox{otherwise}
\end{array}
\right.
 - 2 I_{nz} I_{mz} \right],~~~ \\
b_{nm}&=&-\frac{1}{4}\gamma_{n}\gamma_{m}\hbar\frac{1 - 3 \cos^{2}{\theta_{nm}}}{R^{3}_{nm}}.
\end{eqnarray}
Here, $I_{n\pm}$ denote the raising and lowering operator for the
$n$th nuclear spin and $\gamma_{n}$ its gyromagnetic
ratio, $R_{nm}$ denotes the distance between nucleus $n$ and $m$, and
$\theta_{nm}$ the angle of the vector between them relative to the
applied magnetic field.
Flip-flop interactions between unlike spins are suppressed via energy
conservation due to the applied magnetic field.

The qubit-bath interactions are also due to dipolar
coupling.  (We refer to the central spin as a qubit to distinguish it
from the nuclear spin bath.)  
However, in this instance we disregard the flip-flop
interactions which are suppressed via energy conservation as a result
of the applied magnetic field and 
differing gyromagnetic ratios between the qubit and spins in the
bath.  Instead, we consider just the $I_{nz} I_{mz}$ term which will 
contribute to dephasing:
\begin{eqnarray}
\label{H_A}
{\cal H}_A &\approx& \sum_n A_n I_{nz} S_{z}, \\
A_n &=& \gamma_{\mbox{\tiny D}}\gamma_{n}\hbar\frac{1 - 3
  \cos^{2}{\theta_{n}}}{R^{3}_{n}},
\end{eqnarray}
where $S_z$ is a nuclear spin operator for the P donor nucleus,
$\gamma_{\mbox{\tiny D}}$ is the gyromagnetic ratio of the donor
nucleus, $R_n$ is the
distance of nucleus $n$ for the P donor, and $\theta_n$ is the angle
of the vector from the P donor to nucleus $n$ relative to the applied
magnetic field.  
Note that the $I_{nz} I_{mz}$ term of Eq.~(\ref{H_B})
requires an extra factor of $1/2$ to account for interchanging $n$ and
$m$ that is not required in Eq.~(\ref{H_A}).

The free evolution Hamiltonian is given by ${\cal H}_A + {\cal H}_B$.
We represent the Hahn echo sequence by
$\tau\rightarrow\pi\rightarrow\tau$ 
which we take to mean: free evolution for a time $\tau$ followed by an NMR
$\pi$-pulse rotation of the P donor nuclear spin
about an axis perpendicular to the magnetic
field, and then free evolution for another $\tau$ in time.  The total 
time for this pulse sequence is $t = 2 \tau$.
In order to measure the decoherence of the P donor nuclear
spin, we compute the magnitude of the expectation value of a P donor spin
that is initialized perpendicular to the magnetic field 
(for maximal spectral diffusion) and evolves
according to the Hahn echo sequence.
When normalized to a maximum value of one, this expectation value, as
a function of $\tau$, defines the Hahn echo envelope.
Where $\langle...\rangle$ is the quantum mechanical average over
initial bath states, this Hahn echo envelope may be expressed as\cite{witzelHahn}
\begin{equation}
\label{v_Hahn}
v_{\mbox{\tiny Hahn}}(\tau)=
\left\langle
\left[U_{\mbox{\tiny Hahn}}^{-}(\tau)\right]^{\dag} 
U_{\mbox{\tiny Hahn}}^{+}(\tau) \right\rangle,
\end{equation}
where $U_{\mbox{\tiny Hahn}}^{\pm}(\tau)$ are nuclear evolution
operators in the context of a Hahn echo sequence with
an initially up ($+$) or down ($-$) electron spin such that
\begin{eqnarray}
\label{UHahnpm}
U_{\mbox{\tiny Hahn}}^{\pm}(\tau) &=& 
U^{\mp}_0(\tau) U^{\pm}_0(\tau), \\
\label{U0pm}
U^{\pm}_0(\tau)&=&\textrm{e}^{-\imath{\cal H}_{\pm} t}, \\
\label{Hpm} 
{\cal H}_{\pm}&=&{\cal H}_{B} \pm \frac{1}{2} \sum_{n} A_{n} I_{nz},
\end{eqnarray}
assuming ideal $\pi$-pulses.
Our calculations assume complete disorder of the initial nuclear spin
bath states, expected in the large nuclear temperature limit
($T \gg mK$) without dynamic nuclear polarization or some such
technique to generate order; the quantum mechanical average,
$\langle...\rangle$, of Eq.~(\ref{v_Hahn}) then simply
averages
over all
possible nuclear states with equal weighting (that is, the trace of
the nuclear spin operators divided by the number of nuclear spins states).

Using the formalism that we have laid out in Ref.~\onlinecite{witzelHahn}, we
can expand Eq.~(\ref{v_Hahn}) in a way that includes contributions
from ``clusters'' of successively increasing size.
Defining $v_{\cal C}(\tau)$ as the solution to 
$v_{\mbox{\tiny Hahn}}(\tau)$ [Eq.~(\ref{v_Hahn})] when only
considering those nuclei in some cluster (set) ${\cal C}$, 
a ``cluster contribution'' is recursively defined and computed 
by\cite{witzelHahn}
\begin{equation}
\label{vC'_recursive}
v_{\cal C}'(\tau) = v_{\cal C}(\tau) -
\sum_{
\substack{
\left\{{\cal C}_i\right\}~\mbox{\scriptsize{disjoint}}, \\
{\cal C}_i \ne \emptyset,~{\cal C}_i~\subset~{\cal C}
}}
\prod_i v_{{\cal C}_i}'(\tau),
\end{equation}
subtracting from $v_{\cal C}(\tau)$ 
the sum of all products
of contributions from disjoint sets of clusters contained in 
${\cal C}$.
Defined in this way, a cluster contribution can only be significant 
when interactions between nuclei in the set are significant and 
no part is isolated from the rest.  
We consider only local dipolar interactions in the
current work, and thus contributions only arise when the nuclei in the
set are spatially clustered together; hence we refer to these sets as
clusters.  
The $k$th order of the cluster expansion for the echo [Eq.~(\ref{v_Hahn})] is then\cite{witzelHahn}
\begin{equation}
\label{idealClusterExpansion}
v^{(k)}(\tau) =
\sum_{
\substack{
\left\{{\cal C}_i\right\}~\mbox{\scriptsize{disjoint}}, \\
{\cal C}_i \ne \emptyset,~\lvert{\cal C}_i\rvert \leq k
}}
\prod_i v_{{\cal C}_i}'(\tau),
\end{equation}
and involves clusters of size no larger than $k$.
To simplify the calculation, the logarithm of the echo may be approximated as
\begin{equation}
\label{clusterExpansion}
\ln{\left[v^{(k)}(\tau)\right]} \approx
\sum_{\left|{\cal C}\right| \leq k}  v_{{\cal C}}'(\tau),
\end{equation}
and corrections to this approximation, discussed in
Ref.~\onlinecite{witzelHahn}, may be computed to test its accuracy and
are expected to be $O(1/N)$, where $N$ is the number contributing spins
in the bath.

We have performed such cluster expansion calculations 
to successively
approximate Eq.~(\ref{v_Hahn}) for two different systems.
In both systems, we have a P donor atom with 
$\gamma_{\mbox{\tiny D}} = \gamma_{\mbox{\tiny P}} = 1.08 \times 10^4~\mbox{(s G)$^{-1}$}$, and
we have chosen the applied magnetic field to point along one of the
conventional axes directions (e.g., $B || [001]$).
In our figures, we plot ``memory loss'' versus total echo time ($2
\tau$) where we define
memory loss as one minus the echo envelope, $1 - v_{\mbox{\tiny
    Hahn}}(\tau)$, and we only show results where the cluster
expansion is rapidly convergent and the estimated correction to
Eq.~(\ref{clusterExpansion}) is negligible.
We show Hahn echo results (as well as CPMG results which will be
discussed in the next section) for GaAs:P in
Fig.~\ref{figGaAsP}; in this system, 
$\gamma_{n} = 4.58, 8.16,~\mbox{and}~6.42 \times 10^3 \mbox{(s G)$^{-1}$}$
for $^{75}$As, $^{71}$Ga, and $^{69}$Ga, respectively, and all differ 
from $\gamma_{\mbox{\tiny P}}$.
We additionally show Hahn echo results (and CPMG results) for Si:P in 
Figs.~\ref{figNatSi} and \ref{figPurifiedSi}; 
$\gamma_n = 5.31 \times 10^3~\mbox{(s G)$^{-1}$}$ 
for $^{29}$Si which also differs from $\gamma_{\mbox{\tiny P}}$.
Unlike GaAs, Si has stable isotopes ($^{28}$Si, $^{30}$Si) with zero
spin.  Among its stable isotopes, only $^{29}$Si, which has a natural 
abundance of $4.67\%$ and a spin of $1/2$, has a nonzero spin.
Isotopic purification can reduce the amount of $^{29}$Si and
thereby diminish spectral diffusion caused by the 
nuclear spin bath.
For generality, we define $f$ to be the fraction
of Si that is the $^{29}$Si isotope.
Figure~\ref{figNatSi} shows results in a natural Si bath ($f=0.0467$), while
Fig.~\ref{figPurifiedSi} shows, for comparison, results in a bath of
Si isotopically purified to $f=0.01$.

As an alternative to the cluster expansion, a lowest order expansion of 
$v_{\mbox{\tiny Hahn}}(\tau)$ [Eq.~(\ref{v_Hahn})] in $\tau$ reveals that 
$v_{\mbox{\tiny Hahn}}(\tau) = 1 - O(\tau^4)$, and this simple lowest
order approximation compares favorably with our cluster expansion
results where $v_{\mbox{\tiny Hahn}}(\tau) \ll 1$.  We can be slightly
more sophisticated with this small $\tau$ approximation by applying it
within the context of the cluster expansion; noting that $v_{\cal
  C}'(\tau) = O(\tau^4)$, Eq.~(\ref{clusterExpansion}) implies that
$v_{\mbox{\tiny Hahn}}(\tau) = \exp{\left[-O(\tau^4\right)]}$.
As discussed in previous work,\cite{witzelHahn}
convergence for large systems (i.e., a large bath) is 
attained in this way by embedding the $\tau$
expansion, which is not necessarily convergent for large systems, within
the context of the cluster expansion.

In the lowest order $\tau$ approximation discussed above, 
the GaAs:P result becomes
\begin{equation}
\label{GaAsP_Hahn_approx}
\ln{\left[v_{\mbox{\tiny Hahn}}(\tau)\right]} \approx
-\left(\frac{\tau}{260~\mbox{$\mu$s}}\right)^{4}.
\end{equation}
The exact (convergent) results plotted in Fig.~\ref{figGaAsP}
do not visibly differ from Eq.~(\ref{GaAsP_Hahn_approx}); therefore
this time expansion approximation is valid in the region in which the
cluster expansion converges.  By applying this lowest order time
expansion approximation, as well as the lowest order in the cluster
expansion itself (involving only pair contributions), the Si:P result becomes
\begin{equation}
\label{SiP_Hahn_approx}
\ln{\left[v_{\mbox{\tiny Hahn}}(\tau)\right]} \approx
- f^2 \left(\frac{\tau}{1.05~\mbox{ms}}\right)^{4}. 
\end{equation}
The $f^2$ dependence simply arises from the fact that, in this
approximation, all contributions are from pairs of nuclei.  Dotted
lines in Figs.~\ref{figNatSi} and \ref{figPurifiedSi} show the lowest
order approximation of Eq.~(\ref{SiP_Hahn_approx}) for the Hahn echo (as
well as the lowest order approximation for the CPMG pulse sequence which
will be discussed in the next section).  The exact (convergent)
results exhibit a slight disagreement with the lowest order
approximation as the cluster expansion nears the point of its divergence.

In previous studies
of electron spin dephasing\cite{witzelHahn, witzelCPMG} we noted good
convergence of the cluster expansion for the full decay of the echoes 
as a result of a particularly
applicable intrabath perturbation that treats the coupling between
members of the bath as a perturbation in the Hamiltonian.
In that situation, the hyperfine coupling between
the electron and the nuclei was much stronger than the dipolar coupling among
nuclei.  In the case of nuclear spin dephasing, however, the coupling
to our central nucleus is no stronger than the coupling between
members of the bath.  In fact, the effective smallness of $\tau$
(compared to the inverse energies of the problem) is apparently the
only reason for any cluster expansion convergence in the current study.
As a consequence, the cluster expansion
convergence is only attainable for short times (small values of
$\tau$) as indicated in
Figs.~\ref{figGaAsP}-\ref{figPurifiedSi}.
However, we do obtain convergent and accurate results of the
{\it initial} quantum memory loss in the duration of high fidelity qubit
memory retention which is of greatest significance in the context of
quantum computing.

\section{Multiple Pulse Refocusing}
\label{CPMG}
It is well known in the NMR community that multiple pulse sequences
may be used to prolong qubit coherence for a much greater time than
the simple Hahn echo sequence.  The simplest of these multiple pulse
sequences is known as the
Carr-Purcell-Meiboom-Gill (CPMG)\cite{footnoteCarrPurcell} scheme
in which the Hahn echo
sequence is simply repeated: $(\tau \rightarrow \pi \rightarrow
\tau)^n$.  
In our previous
work\cite{witzelCPMG} on electron spin coherence in semiconductors, 
independently confirmed in Ref.~\onlinecite{yao}, 
we reported that low-order symmetry-related
cancellations occur when an even number of CPMG pulses are applied;
this leads to significantly longer coherence times of even-pulse CPMG
echoes compared with the Hahn echo.  Also, increasing
the number of these pulses will increase the overall coherence time with
the assumption that the applied pulses are ideal.  Consistent with these
previous findings, we find that donor nucleus coherence is significantly
enhanced by applying a two-pulse CPMG relative to the single
pulse Hahn refocusing, and further enhanced, in overall
coherence time, by applying a four-pulse CPMG sequence.
We now consider the application of multiple pulse echo sequences in
prolonging nuclear spin quantum memory in semiconductor quantum
computer architectures.

If we apply $2 \nu$ CPMG pulses, $(\tau \rightarrow \pi \rightarrow
\tau)^{2 \nu}$, the total echo time for this is $t = 4 \nu \tau$, and
the exact echo envelope may be expressed [analogous to
  Eq.~(\ref{v_Hahn})] as
\cite{witzelCPMG}
\begin{equation}
\label{v_CPMG}
v_{\mbox{\tiny CPMG}}(\nu, \tau)=
\left\langle
\left[U_{\mbox{\tiny CPMG}}^{-}(\nu, \tau)\right]^{\dag} 
U_{\mbox{\tiny CPMG}}^{+}(\nu, \tau) \right\rangle,
\end{equation}
where $U_{\mbox{\tiny CPMG}}^{\pm}(\nu, \tau)$ are nuclear evolution
operators in the context of a CPMG echo sequence with
an initially up ($+$) or down ($-$) electron spin such that
\begin{eqnarray}
\label{UCPMGpm}
U_{\mbox{\tiny CPMG}}^{\pm}(\nu, \tau) &=& 
\left[U^{\pm}_0(\tau) U^{\mp}_0(2 \tau) U^{\pm}_0(\tau)\right]^{\nu},
\end{eqnarray}
referring to the definition of $U^{\pm}_0$ 
given in Eq.~(\ref{U0pm}).
The angle brackets, $\langle...\rangle$, of Eq.~(\ref{v_CPMG}) represent
a quantum mechanical average over initial bath states that we 
take to be in a uniform distribution representing a completely
disordered bath as we did in Sec.~\ref{SD_nuc_mem}.

\begin{figure}
\includegraphics[width=3in]{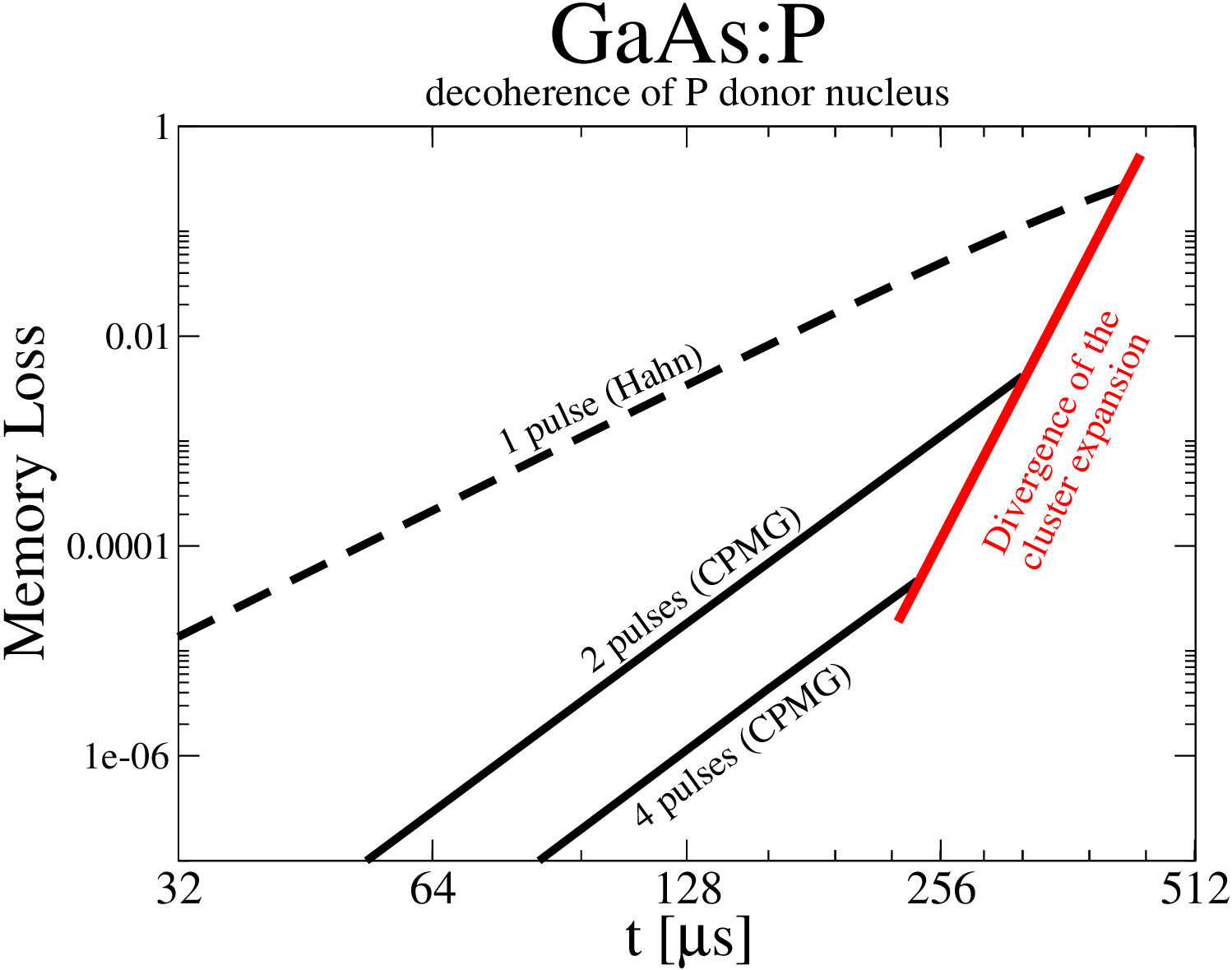}
\caption{
Numerical results of nuclear spin quantum memory loss
for a $^{31}$P donor nucleus that replaces an As atom in bulk GaAs.  
We define memory loss as one minus the normalized echo and plot this in a
log-log scale as a function of the total echo time.  The dashed line gives
the Hahn echo results and the solid lines give CPMG echo results for
two and four pulses.  At some point for each type of echo sequence,
the cluster expansion fails to converge.
\label{figGaAsP}}
\end{figure}

\begin{figure}
\includegraphics[width=3in]{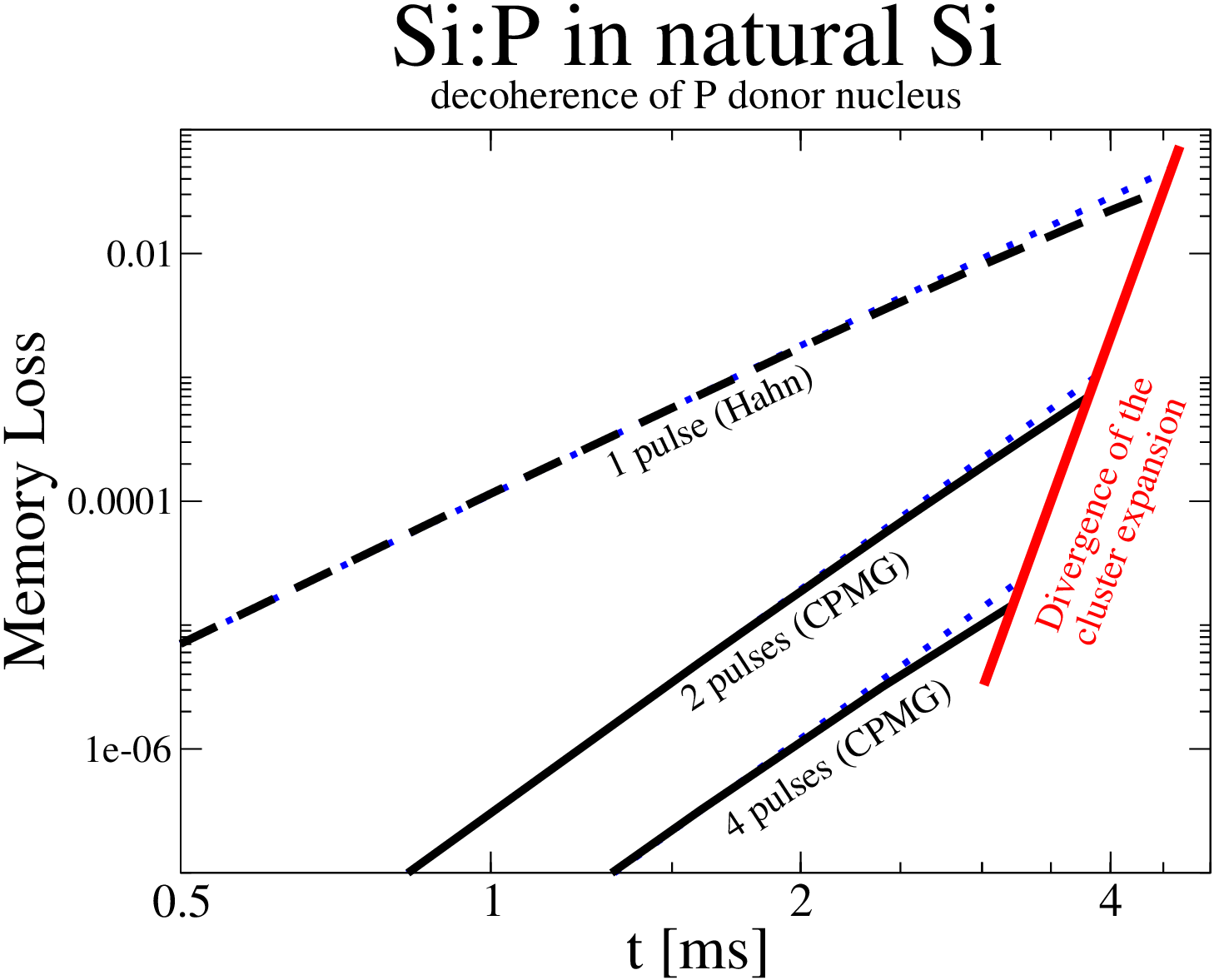}
\caption{
Numerical results of nuclear spin quantum memory loss
for a $^{31}$P donor nucleus in bulk Si.  
We define memory
loss as one minus the normalized echo and plot this in a
log-log scale as a function of the total echo time.  
The dashed line gives
the Hahn echo results and the solid lines give CPMG echo results for
two and four pulses.  Dotted lines give corresponding results, for comparison, 
obtained from the lowest order expansions 
provided by Eqs.~(\ref{SiP_Hahn_approx}) and (\ref{SiP_CPMG_approx}).
At some point for each type of echo sequence,
the cluster expansion fails to converge.
\label{figNatSi}}
\end{figure}

\begin{figure}
\includegraphics[width=3in]{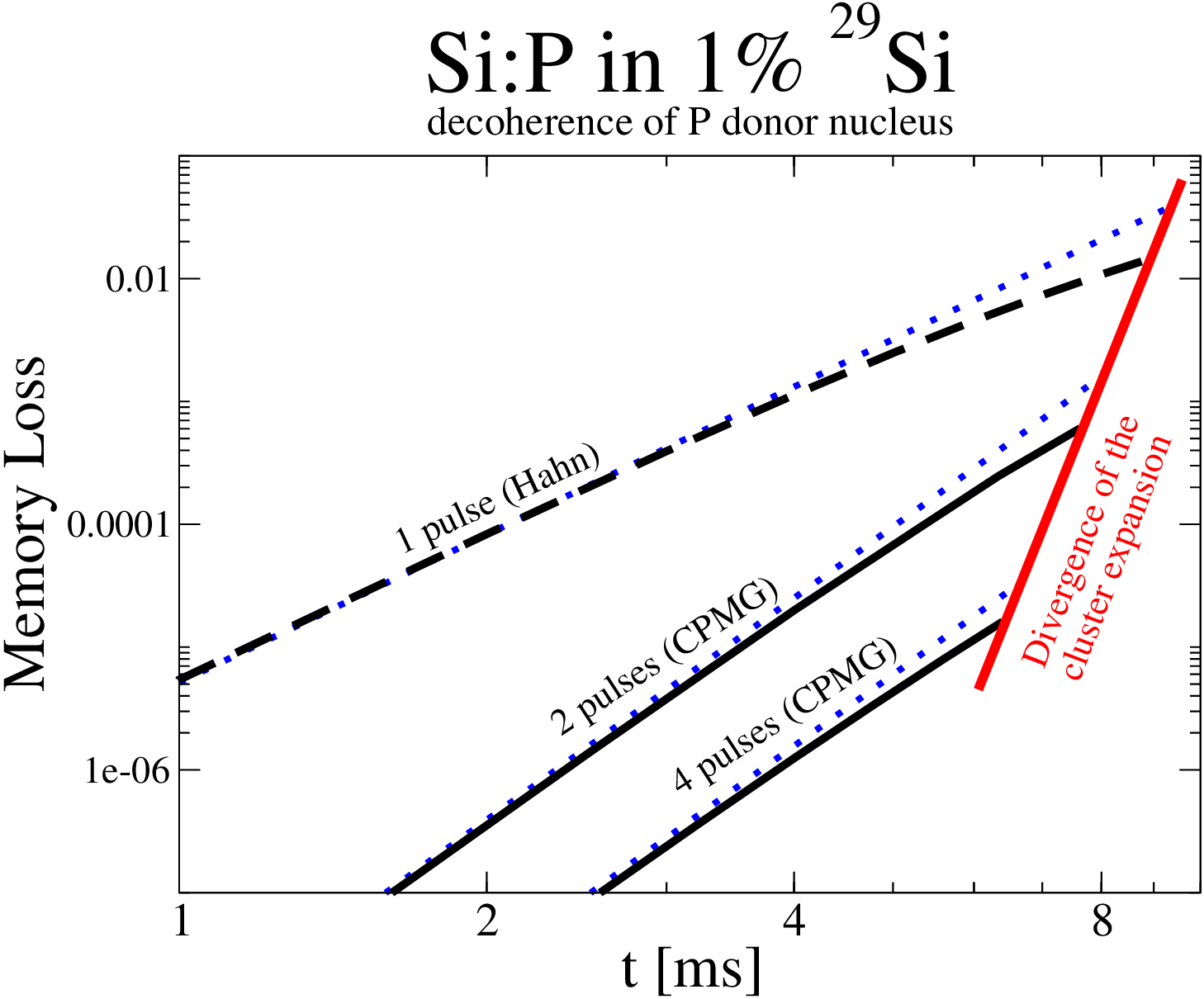}
\caption{
Equivalent to Fig.~\ref{figNatSi} except that results are shown for Si
purified to $1\%$ $^{29}$Si.  Lowest order results given by
Eqs.~(\ref{SiP_Hahn_approx}) and (\ref{SiP_CPMG_approx}) are shown by
the dotted lines.
Isotopic purification enhances coherence
as predicted in these equations.
\label{figPurifiedSi}}
\end{figure}

The cluster expansion is performed as described in the
previous section using Eqs.~(\ref{vC'_recursive}) and
(\ref{clusterExpansion}) with the only exception that 
$v_{\cal C}(\tau)$ should now represent the solution,
while considering only the nuclei in cluster ${\cal C}$,
to Eq.~(\ref{v_CPMG}).
We have performed these cluster expansion calculations to successively
approximate Eq.~(\ref{v_CPMG}) for the GaAs:P and Si:P systems.
We again plot ``memory loss'' versus total echo time ($4 \nu \tau$) 
where memory loss is defined as $1 - v_{\mbox{\tiny CPMG}}(\tau)$.
Results for $2$ and $4$ pulses ($\nu = 1$ and $2$, respectively) in the
GaAs:P system are shown in Fig.~\ref{figGaAsP} where
they are compared with Hahn echo results.  Similarly, results for the
natural and isotopically purified Si:P systems are, respectively, shown
in Figs.~\ref{figNatSi} and \ref{figPurifiedSi} where they are also
compared with corresponding Hahn echo results.  A spin coherence
enhancement when applying two pulses versus one pulse (Hahn echo) is
apparent in the plots for each of these systems.

As shown in Ref.~\onlinecite{witzelCPMG}, if we apply a lowest order 
time expansion {\it and} an intrabath perturbation (with coupling
between nuclei as a perturbation in the Hamiltonian)
approximation in the cluster
contribution terms that go into the cluster expansion calculation, we
find that $v_{\mbox{\tiny CPMG}}(\tau) \sim
\exp{\left[-O(\nu^2 \tau^6\right)]} = \exp{\left[-O(\nu^{-4} t^6\right)]}$.
In this lowest order approximation, the general GaAs:P result becomes
\begin{eqnarray}
\nonumber
\ln{\left[v_{\mbox{\tiny CPMG}}(\tau)\right]} &\approx&
-\nu^2 \left(\frac{\tau}{195~\mbox{$\mu$s}}\right)^{6}\\
\label{GaAsP_CPMG_approx}
 &=&
-\nu^{-4} \left(\frac{t}{780~\mbox{$\mu$s}}\right)^{6},
\end{eqnarray}
as compared with
$\ln{\left[v_{\mbox{\tiny Hahn}}(\tau)\right]} \approx
-\left(t / 520~\mbox{$\mu$s}\right)^{4}$
[Eq.~(\ref{GaAsP_Hahn_approx}) rewritten in terms of $t$].
The exact (convergent) results plotted in Fig.~\ref{figGaAsP}
do not visibly differ from Eq.~(\ref{GaAsP_CPMG_approx}); therefore
this approximation is valid in the region in which the
cluster expansion converges.
This short time approximation equation
may serve as a useful educated guess (estimate) at
times beyond cluster expansion convergence although the decay is
expected to eventually exhibit exponential, Markovian behavior.  
If we do extrapolate Eq.~(\ref{GaAsP_CPMG_approx}) and
define $T_2$ as the time in which the extrapolated echo reaches
$1/e$, then we have $T_2 = \nu^{0.67} \times 780~\mbox{$\mu$s}$ for
even CPMG echoes.  This gives a factor of $6.5$ increase of nuclear
spin coherence times relative to
the electron spin quantum dot coherence times reported in 
Ref.~\onlinecite{witzelCPMG}.

In the range of cluster convergence, where we have confidence in the
accuracy of our results for the model that we have used, 
we observe high fidelity memory retention with a low loss of $10^{-6}$ up to
$80-120~\mbox{$\mu$s}$ for two
or four-pulse CPMG sequences.  Increasing the number of CPMG pulses
can improve fidelity versus time theoretically; however, in practice
one will be limited by the finite time required to perform each
$\pi$-pulse and by the accumulation of errors resulting from multiple
pulses.  Because our analysis assumes that each pulse is ideal, 
which is unlikely in practice, these theoretical results for $T_2$
should thus be regarded as an upper bound on the nuclear memory time.

Applying the lowest order approximation to the Si:P system, we
similarly obtain
\begin{eqnarray}
\nonumber
\ln{\left[v_{\mbox{\tiny CPMG}}(\tau)\right]} &\approx&
-\nu^2 f^2 \left[  \left(\frac{\tau}{1.35~\mbox{ms}}\right)^{6} 
+ f \left(\frac{\tau}{0.70~\mbox{ms}}\right)^{6} \right] \\
\nonumber
&=&
-\nu^{-4} f^2 \left(\frac{t}{5.4~\mbox{ms}}\right)^{6} \\
\label{SiP_CPMG_approx}
&&-\nu^{-4} f^3 \left(\frac{t}{2.8~\mbox{ms}}\right)^{6}.
\end{eqnarray}
as compared with
$\ln{\left[v_{\mbox{\tiny Hahn}}(\tau)\right]} \approx
- f^2 \left(t / 2.1~\mbox{ms}\right)^{4}$
[Eq.~(\ref{SiP_Hahn_approx}) rewritten in terms of $t$].
These lowest order results are shown as dotted
lines in Figs.~\ref{figNatSi} and \ref{figPurifiedSi}.
The exact (convergent) results initially match this lowest order
approximation, but show a slight disagreement as 
the cluster expansion nears the point of its divergence.

As with Eqs.~(\ref{GaAsP_Hahn_approx}) and (\ref{GaAsP_CPMG_approx}),
the above equations may serve as a useful educated guess (estimate) at
times beyond cluster expansion convergence.  
Initially, at least, Figs.~\ref{figNatSi}
and \ref{figPurifiedSi} show that Eqs.~(\ref{SiP_Hahn_approx}) and
(\ref{SiP_CPMG_approx}) provide conservative estimates.
If we do extrapolate these equations and
define $T_2$ as the time in which the extrapolated echo reaches
$1/e$, then we have, for natural Si, $T_2 = \nu^{0.67} \times 12~\mbox{ms}$ for
even CPMG echoes.  For a small number of pulses, $\nu \sim 1$, this
gives about a factor of $5$ increase of nuclear spin coherence times relative to
the electron spin quantum dot coherence times reported in 
Ref.~\onlinecite{witzelCPMG}; this comparison factor increases as we
increase the number of pulses because electron spin
decay time\cite{witzelCPMG} scales with a smaller power of $\nu$ ($\nu^{0.53}$).
In the range of cluster convergence, where we have confidence in the
accuracy of our results for the model that we have used, 
we observe high fidelity memory retention with a low loss of $10^{-6}$ up to
$1-2~\mbox{ms}$ for two
or four-pulse CPMG sequences in natural Si and nearly up to
$4~\mbox{ms}$ for $1\%$ purified Si.

By implementing CPMG pulse sequences with just a few even number of pulses,
high fidelity (with loss below $10^{-6}$) qubit retention times are
theoretically observed on the order of 
$100~\mbox{$\mu$s}$ for GaAs systems and on the order of
milliseconds for Si:P systems.
We emphasize that although we are unable to achieve convergence beyond
the initial decay (for reasons discussed at the end of
Sec.~\ref{SD_nuc_mem}) which affects the accuracy of our extrapolated
estimate for $T_2$, itself, we accurately estimate the initial-time
coherent memory loss (i.e., the loss of the first
$10^{-4}-10^{-6}$ fraction of coherence) which is the most important
ingredient for quantum computation considerations.

\section{Intrinsic Spin Relaxation}
\label{nuc_transport}

In addition to storing quantum information on individual
nuclear spins, there is also a proposal\cite{Taylor03} to store
quantum information in coherent states of nuclear spin ensembles.
The proposal suggests transferring the spin state of a quantum dot
electron onto coherent spin states of the nuclei within the dot via
resonance.  Before the information is transferred, the nuclei must be
polarized or cooled into
a so-called dark state that will only interact with electrons of one type of 
polarization and not the other; this can be accomplished
by injecting polarized electrons into the dot.\cite{ImamogluPRL03}
Then through the hyperfine
interactions between the electron and the nuclei, one may induce
Rabi oscillations via resonant tuning of the external magnetic field 
that will pass the quantum information back and forth between the
electron spin and coherent states of the nuclear ensemble.
In this way, one can transfer the quantum information from the
electron spin to the nuclear spins.  
The electron can then be moved off of the dot, and the
nuclear ensemble will store the quantum state for
milliseconds\cite{Deng05} until it is retrieved back onto an electron
spin state by the same process using Rabi oscillations.

The coherent state of the nuclear spin ensemble in such a proposal
can be destroyed by nuclear spin dynamics induced, for example, by dipolar
interactions between nuclear spins.  
If an electron is kept on
the dot during the storage period, differences of hyperfine energies
will suppress dipolar interactions that lead to decoherence; however,
these differences of hyperfine energies in themselves result in an
even more rapid loss of coherence due to developing phase differences
of the nuclear spins.\cite{Deng05}  Thus it is best to move the
electron off of the dot; then the quantum memory time for this
proposal will be determined solely by interactions (e.g., dipolar) among the nuclei.
Assuming that the quantum state is imprinted on a nuclear spin
ensemble that is fully polarized initially, the stored nuclear state
will be a superpositon of the fully polarized state and each possible
state with only one spin oppositely polarized; then the coherent state
is lost, in particular, as the oppositely polarized nucleus, of the various
superposition states, is transported off of the dot by dipolar
flip-flop processes.
This situation is analyzed for specific quantum dot geometries in
Ref.~\onlinecite{Deng05}.

We will study the problem of the decoherence of such a nuclear spin
ensemble state in a more general and simple way that
provides a lower bound limit of coherence time (in the ideal case).  Instead of
considering a specific dot geometry and studying decoherence as a
process in which spins are transported off of the dot, we simply
consider the spin relaxation of a single nuclear spin in a bath of fully
polarized spins.  This provides a lower bound, or pessimistic
estimate, of the coherence time of an ensemble nuclear spin state such
as the previous proposal and gives a general consideration for similar
proposals that may arise.
This simple analysis also gives a nice theoretical counterpart to 
spectral diffusion which is not possible in
this scenario; in the limit of a strong applied magnetic field 
flip-flop processes cannot occur among spins of the polarized bath.

With this motivation, we study a well-defined problem to compute
the lifetime of quantum information stored in a central ``intrinsic''
(as opposed to a donor) nuclear spin, surrounded by a bath of fully
polarized nuclei of the same type, 
and only considering nuclear spin degrees of freedom
in the limit of a strong applied magnetic field.
The precise scenario being considered here in the context of nuclear
spin quantum memory lifetime is the following: The central spin in a
polarized nuclear spin bath has its spin opposite to all the other
spins in the bath initially and we are interested in finding out how
this local spin excitation moves (i.e., ``diffuses,'' although
the process is technically not a diffusive process) away from the
central spin due to dipolar coupling with surrounding nuclei.
The central spin serves as the quantum memory
which ``decays'' in time as the state evolves into 
superpositions of excitation states involving spins in the bath.
The effective Hamiltonian for the dipolar interaction, under the influence of 
a strong magnetic field which effectively conserves polarization via
energy conservation, is given by
\begin{equation}
\label{H}
{\cal H} = \sum_{n \ne m} b_{nm} (I_{n+} I_{m-} - 2 I_{nz} I_{mz}),
\end{equation}
where the $I_{n}$ operators operate on the nuclear spin at site $n$ in
the basis in which $z$ is in the direction of the applied magnetic field.
The dipolar coupling is given by
\begin{equation}
\label{b_nm}
b_{nm}=-\frac{1}{4}\gamma_{I}^{2}\hbar^2 \frac{1 - 3 \cos^{2}{\theta_{nm}}}{R^{3}_{nm}},
\end{equation}
where $\theta_{nm}$ is the angle formed between the applied magnetic
field and the bond vector linking the two spins.  In a
uniform applied magnetic field, which we assume, the total Zeeman
energy will be constant when polarization is conserved.
Therefore the magnetic field does not affect the dynamics beyond
determining the preferred $z$ direction and justifying the polarization
conservation approximation of Eq.~(\ref{H}).

In our initial state, we will assume a fully polarized nuclear spin
bath.  We denote the fully polarized state, including the
polarization of the central spin, as $\left\lvert\uparrow^N\right\rangle$.  We
denote the state with the spin of nucleus $n$ lowered once (via the
lowering spin operator) from its fully
polarized state, but the remaining nuclei
polarized, as $\left\lvert n\right\rangle = I_{n-} \left\lvert\uparrow^N\right\rangle$.
We can store a qubit in the superposition of states
$\left\lvert\uparrow^N\right\rangle$ and $\left\lvert 0\right\rangle$.
The state $\left\lvert\uparrow^N\right\rangle$ is an eigenstate of 
Eq.~(\ref{H}) and is therefore stable.  The
$\left\lvert 0\right\rangle$ state, however, can decay as a result of a
flip-flop with some neighboring nucleus; in this section, we will
determine the lifetime of the $\left\lvert 0\right\rangle$ state for
two different types of materials, GaAs and Si, and characterize the
decay curve.  We can solve this problem exactly by using a Hilbert space
composed of the $\left\lvert n\right\rangle$ states as a basis (since
our Hamiltonian does not include any interactions that can take us out
of this state space).
Conveniently, this state space grows linearly with the number of
nuclei (rather than the exponential growth of the Hilbert space that
occurs in general) which allows this problem to be solved exactly
even when there are many neighbors that influence the decay.

We wish to directly compute the probability that the system remains in state
$\left\lvert 0\right\rangle$ after allowing the system to freely evolve
for a time $t$:
\begin{equation}
P_0(t) = \left\lVert \langle 0 \vert \exp{(-i {\cal H} t)} \vert 0
  \rangle \right\rVert^2.
\end{equation}
To solve this problem directly, we first compute the matrix elements
of ${\cal H}$ in the $\left\lvert n\right\rangle$ basis and then we
can exponentiate this matrix via direct diagonalization.  The
off-diagonal elements of ${\cal H}$ are
\begin{equation}
\left<n\right|{\cal H}\left|m\right> = 2 I b_{nm}\mbox{, $\forall~n \ne m$},
\end{equation}
where $I$ is the nuclear spin magnitude, assumed to be the same for
all involved nuclei.  To obtain this we have used
\begin{equation}
\label{Ipm}
I_{\pm} \left|I, m\right> = \sqrt{I (I+1) - m (m \pm 1)} \left|I, m \pm 1\right>,
\end{equation}
where only $I_+ \left|I, I-1\right>$ and $I_- \left|I, I\right>$ are relevant.
We also have
\begin{eqnarray}
\left<n\right|{\cal H}\left|n\right> &=& -2 \sum_{k \ne m} b_{km} I (I - \delta_{nk} -
\delta_{nm}) \\
\label{diag_H}
&=& -2 \sum_{k \ne m} b_{km} I^2 + 4 \sum_{m \ne n} b_{nm} I.
\end{eqnarray}
The first term in Eq. (\ref {diag_H}) is a constant contribution to
the diagonal and therefore a constant, commuting contribution to the 
Hamiltonian that will not affect probabilities.  We can therefore use
an effective Hamiltonian, ${\cal H}'$, in which
\begin{eqnarray}
\left<n\right|{\cal H}'\left|m\right> &=& 2 I b_{nm}\mbox{, $\forall~n \ne m$}, \\
\label{diag_H'}
\left<n\right|{\cal H}'\left|n\right> &=& 4 I \sum_{n \ne m} b_{nm}.
\end{eqnarray}

We may now solve the problem by
successively including increasingly distant neighbors of the central
spin into the simulated system until convergence is achieved.  
In a Bravais lattice with only one type of nucleus,
$\left<n\right|{\cal H}'\left|n\right>$ 
[Eq.~(\ref{diag_H'})]
is, by symmetry, a constant for all $n$ and is therefore irrelevant (another constant
energy in the system that plays no role in the dynamics).  
In lattices with mixed nuclei, this may
not be completely irrelevant; however, we have performed calculations
on real lattice systems with and without this part of the Hamiltonian 
finding that it has little effect on the results except that it artificially
slows down the rate of
convergence as we include increasingly distant neighbors
(because it does lead to a significant finite size effect).

\begin{figure}
\includegraphics[width=3in]{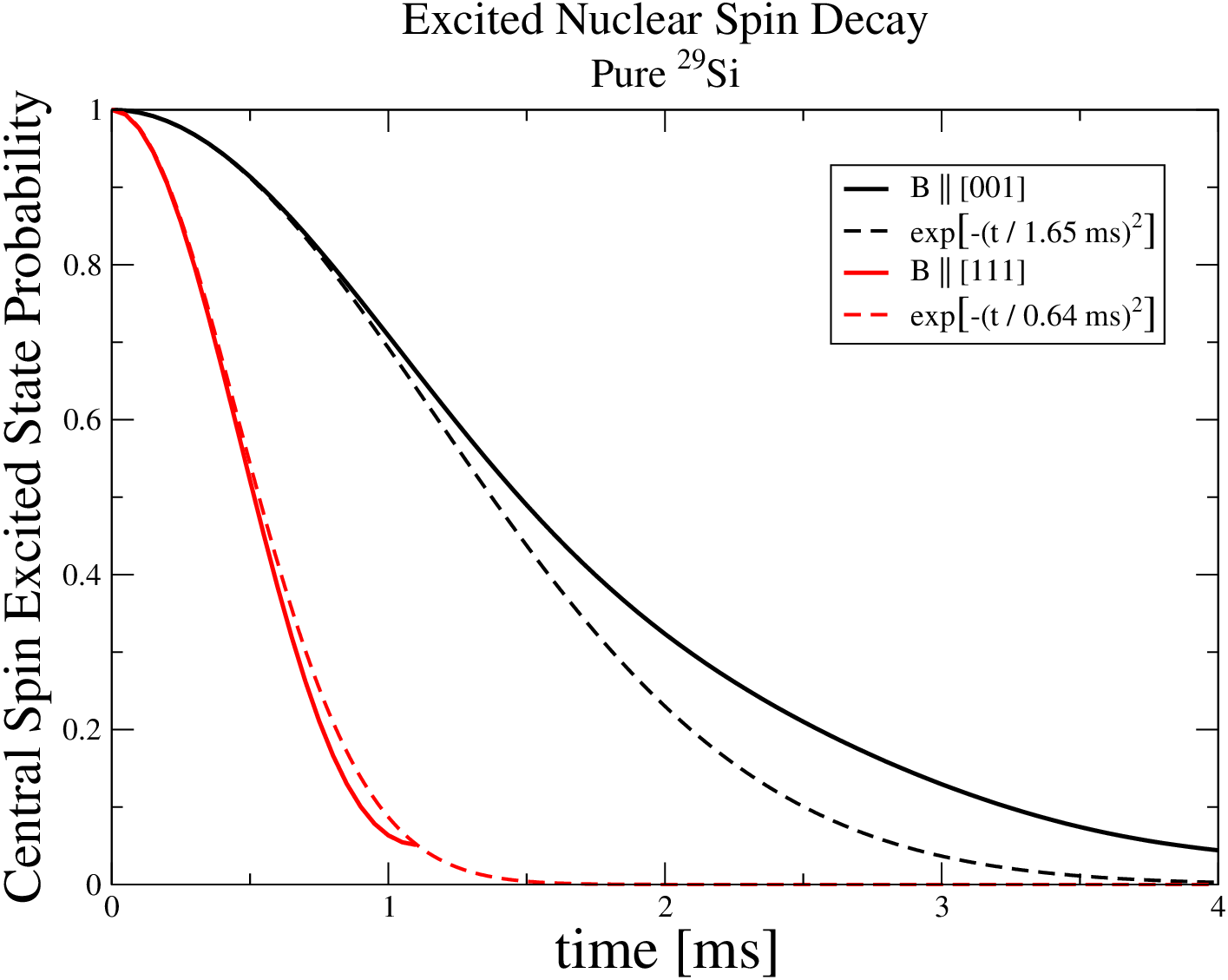}
\caption{
\label{IntrinsicSi100}
Relaxation of a central $^{29}$Si nuclear spin in a bath of
polarized isotopically pure $^{29}$Si nuclei in silicon's diamond lattice
structure in the limit of a strong applied magnetic field along two
directions with extremal results: $B$ along [001] (or, equivalently, along
any of the conventional axes, $x$, $y$, or $z$) and $B$ along [111].
The dashed curves give the corresponding approximate solutions from Eq.~(\ref{P0_approx}).
}
\end{figure}

\begin{figure}
\includegraphics[width=3in]{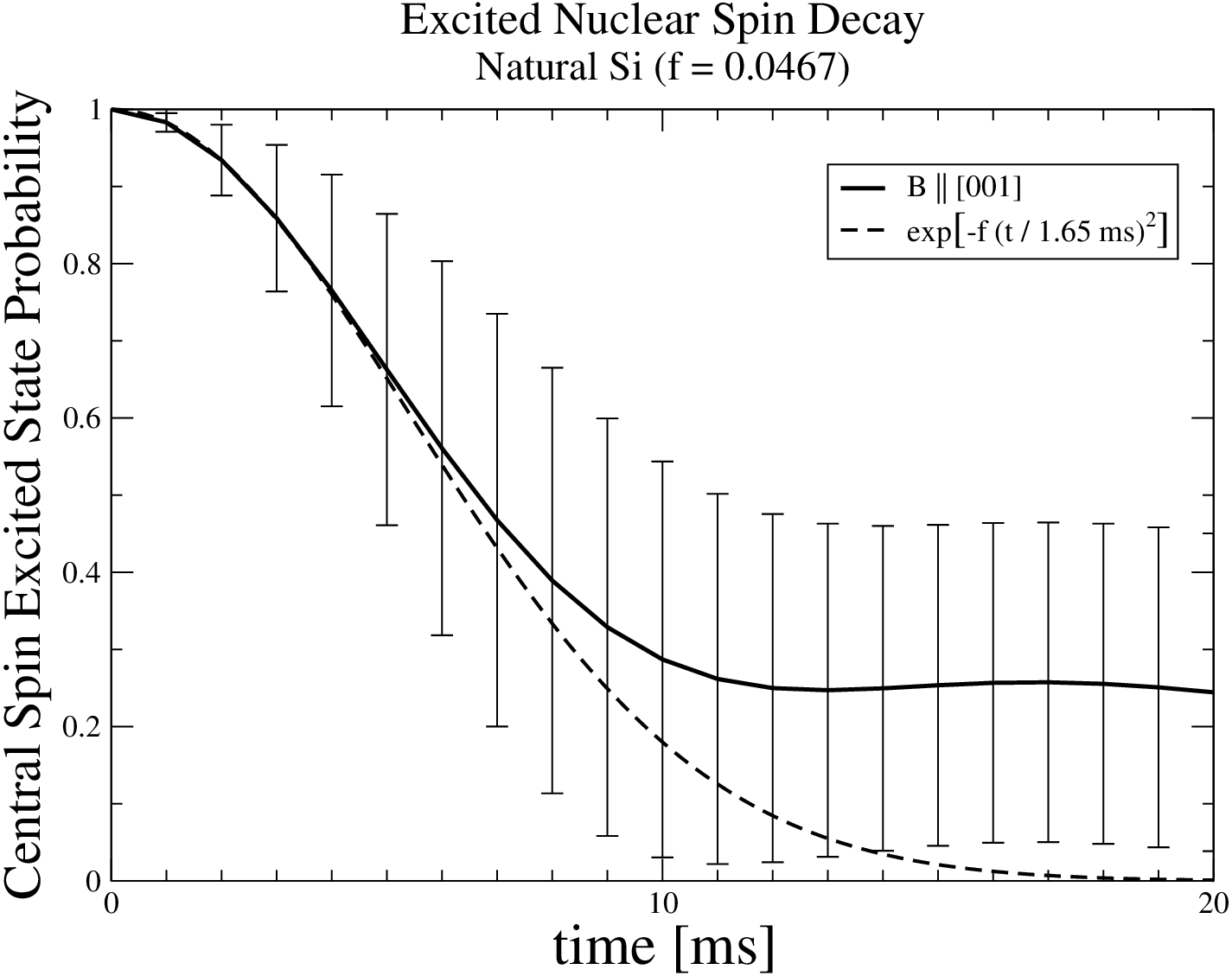}
\caption{
\label{IntrinsicNatSi}
Relaxation of a central $^{29}$Si nuclear spin in a bath of
polarized natural Si nuclei in silicon's diamond lattice structure
in the limit of a strong applied magnetic field along one of the
conventional axes (e.g., [100]).  Error bars indicate the standard
deviation as a result of random isotopic configurations ($4.67\%$
$^{29}$Si and the remaining $^{28}$Si and $^{30}$Si is spinless giving
no contribution to the relaxation).
The dashed curve gives the corresponding approximate solution from Eq.~(\ref{P0_approx}).
}
\end{figure}

Calculated results are shown for pure $^{29}$Si and natural Si in
Figs.~\ref{IntrinsicSi100} and \ref{IntrinsicNatSi},
respectively.  The $^{29}$Si has a spin of $1/2$ and 
$\gamma_I = 5.31 \times 10^3~\mbox{(s G)$^{-1}$}$. 
The other stable isotopes of Si, $^{28}$Si and
$^{30}$Si, have no net nuclear spin and therefore do not contribute to
the decay.  Si has a diamond lattice structure with a lattice
constant of $5.43~\mbox{{\AA}}$.
In natural Si,
there is a $4.67\%$ abundance of $^{29}$Si and the random placement of these
nuclei leads to uncertainty in the results; the standard deviation is
thus shown by the error bars in Fig.~\ref{IntrinsicNatSi}.

\begin{figure}
\includegraphics[width=3in]{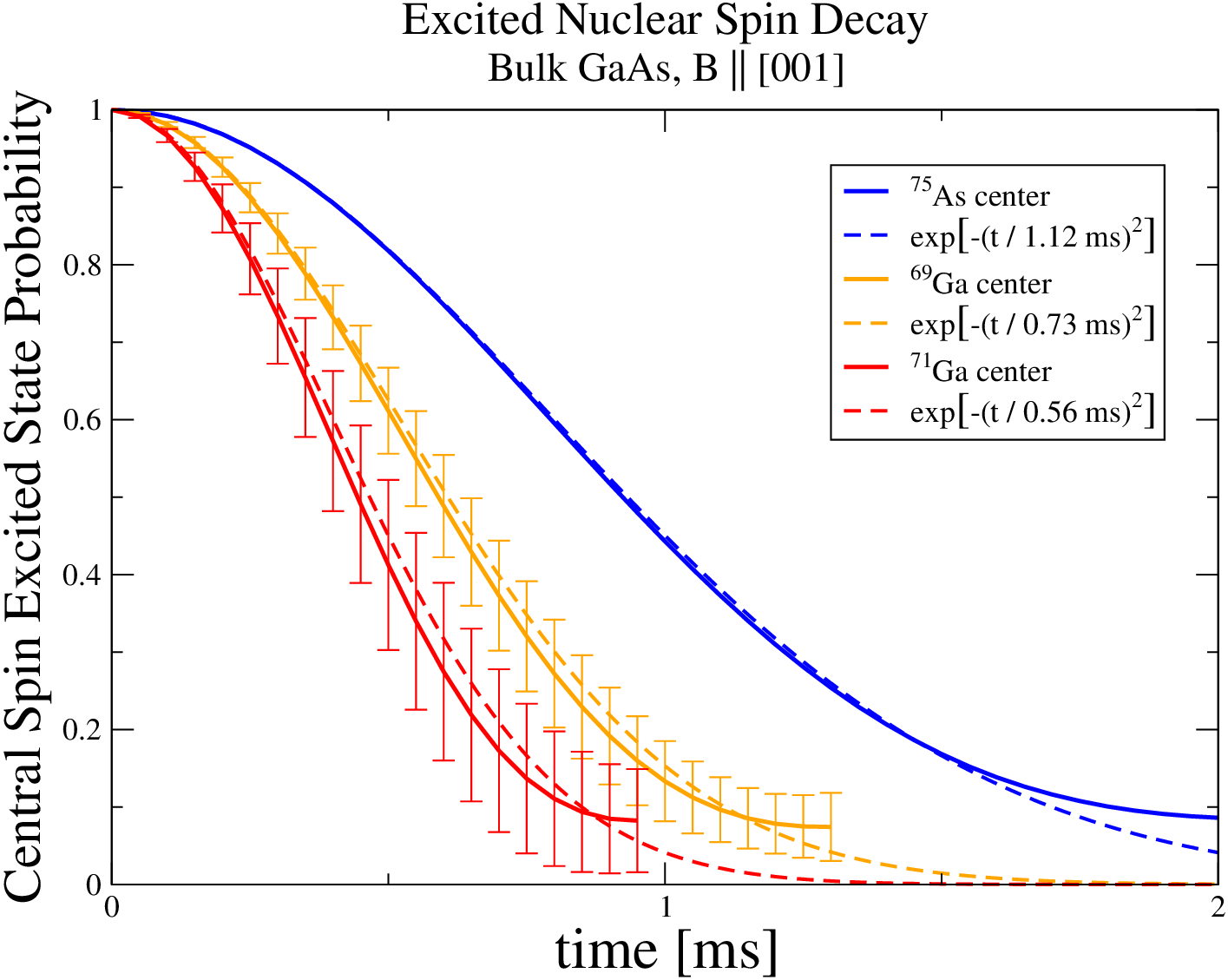}
\caption{
\label{IntrinsicGaAs}
Relaxation of a central $^{75}$As, $^{69}$Ga, or $^{71}$Ga nuclear
spin in a bath of polarized GaAs nuclei in GaAs's zinc-blende lattice structure
in the limit of a strong applied magnetic field along one of the
conventional axes (e.g., [100]).  In this limit, the central spin only
interacts with like nuclei.  The $^{75}$As are arranged with certainty
on one of the fcc lattices, while the $^{69}$Ga and $^{71}$Ga randomly
occupy $60.4\%$ and $39.6\%$ of the other fcc lattice, respectively.
Error bars for the two Ga results indicate the standard
deviation as a result of random isotopic configurations.
The dashed curves give the corresponding approximate solutions from Eq.~(\ref{P0_approx}).
}
\end{figure}

Calculated results are shown for GaAs in
Fig.~\ref{IntrinsicGaAs}.  There are three distinct problems with
the central spin being $^{75}$As, $^{69}$Ga, or $^{71}$Ga which have
respective gyromagnetic ratios of 
$\gamma_I = 4.58,~6.42,~\mbox{and}~8.16 \times 10^3~\mbox{(s G)$^{-1}$}$ and all have
spins of $3/2$.
The GaAs is in a zinc-blende lattice structure with a lattice constant
of $5.65~\mbox{{\AA}}$.  The $^{75}$As occupies
one fcc lattice while the $^{69}$Ga and $^{71}$Ga are randomly
distributed in $60.4\%$ and $39.6\%$ respective concentrations on the
other fcc lattice.  The uncertainty in the random isotopic
configuration of the Ga sublattice is the cause for standard deviations shown
by the error bars in Fig.~\ref{IntrinsicGaAs}.

To approximate the short time behavior, let us first expand to the
lowest order in powers of $t$.  First note that $\langle 0 \vert H'
\vert 0 \rangle = 0$ so there is not a first power of $t$ in the
expansion.  In order to obtain the next order (first nontrivial
order), we have
\begin{eqnarray}
\langle 0 \vert \left({\cal H}'\right)^2 \vert 0 \rangle &=& 
\sum_{n} \langle 0 \vert {\cal H}' \vert n \rangle \langle n \vert
    {\cal H}' \vert 0 \rangle \\
&=& 4 I^2 \sum_{n \ne 0} b_{n0}^2,
\end{eqnarray}
so then
\begin{equation}
\langle 0 \vert \exp{(-i {\cal H} t)} \vert 0 \rangle \approx 1 - 2
I^2 \sum_{n \ne 0} b_{n0}^2 t^2.
\end{equation}
Squaring this and keeping only terms up to $t^2$, we finally obtain
\begin{equation}
P_0 \approx 1 - (2 I)^2 \sum_{n \ne 0} b_{n0}^2 t^2.
\end{equation}

To better match the shape of the decay curves, we use an
exponential in time that is consistent with the above approximation:
\begin{equation}
\label{P0_approx}
P_0 \approx \exp{\left(-f (2 I)^2 \sum_{n \ne 0} b_{n0}^2 t^2\right)}.
\end{equation}
In order to effectively average over isotopic configurations, we
have now explicitly inserted the dependence on the fraction, $f$, of
nuclei in the lattice (or sublattice) that is of the same type as the
central spin.
This equation is indeed a good
approximation to the initial decay and is exhibited by the dashed
curves in Figs.~\ref{IntrinsicSi100},
\ref{IntrinsicNatSi}, and \ref{IntrinsicGaAs}.  A good
characterization of the lifetime is the time in which this
approximation reaches $1/e$ which is given by
\begin{equation}
t_0 = 1/\sqrt{f (2 I)^2 \sum_{n \ne 0} b_{n0}^2}.
\end{equation}
This also indicates how the lifetime scales with isotopic purification
(e.g., in Si).  Using this $t_0$ to define the lifetime, natural Si
has a lifetime of $1.7$~ms, and for GaAs, $^{75}$As, $^{69}$Ga, and
$^{71}$Ga have lifetimes of $1.1$, $0.7$, and $0.6$~ms,
respectively.
Note that while there is a greater abundance of $^{75}$As than the
other nuclear types in GaAs, it has a longer lifetime due to its
relatively small gyromagnetic ratio.

\section{Conclusion}
\label{conclusion}
We have analyzed in this work two complementary theoretical issues
regarding the prospect for using nuclear spins in semiconductor
materials, specifically GaAs and Si (with and without P doping), as 
long-lived quantum memory.  We find in general that solid state nuclear
quantum memory time could be one to two orders of magnitude longer
than the corresponding electron spin coherence times.  Thus using
nuclear spins as quantum memory in semiconductor nanostructures may be
useful in the context of solid state quantum information processing
architectures.

We address two complementary situations of
solid state nuclear spin coherence, both taken in the limit of a
strong applied magnetic field: (1) quantum memory
stored in a donor nuclear spin in a surrounding unpolarized
nuclear spin bath where the main memory loss (i.e., decoherence) occurs
through the spectral diffusion mechanism associated with the
non-Markovian temporally fluctuating random magnetic field created by
the dipolar flip-flops in the nuclear spin bath 
(Secs.~\ref{SD_nuc_mem} and \ref{CPMG}); and (2) quantum memory stored in an
intrinsic spin surrounded by a spin-polarized nuclear bath
where the spin
relaxes through direct flip-flop interactions with neighboring 
nuclei of the same type (Sec.~\ref{nuc_transport}).  
These are mutually exclusive situations; spectral diffusion disregards spin relaxation (justified
in a strong magnetic field), and a polarized bath cannot induce
spectral diffusion.

In case (1), we note that our $T_2$ dephasing times 
define  characteristic time scales for the {\it initial} part of the decay
that is governed by non-Markovian dynamics in the bath.
Our $T_2$ values of $520~\mbox{$\mu$s}$ for P in GaAs and 
$9.7~\mbox{ms}$ for P in natural Si are consistent with the long-time exponential decay
behavior studied in NMR,\cite{NMRlit} but it is important to note that
we expect a different short time echo decay behavior of the form
$\exp{\left[-(t/T_2)^4\right]} = \exp{\left[-(2 \tau/T_2)^4\right]}$ 
for the Hahn sequence. 
This has consequences for the nature of high-fidelity quantum memory
retention that is most important for quantum computing considerations.
Longer memory times may be achieved in case (1) by using 
composite CPMG pulse sequences; with an even number of $2 \nu$ pulses
and total sequence time $t= 4 \nu \tau$,
the echo decay behavior is of the form 
$\exp{\left[-\nu^2 (\tau/\tau_0)^6\right]} = 
\exp{\left[-\nu^{-4} (t/t_0)^6\right]} =
\exp{\left[-t^2 \tau^4 / (16 \tau_0^6)\right]}$.
The latter expression is relevant to a train of pulses with fixed
$\tau$; because we consider the non-Markovian, nonexponential 
regime of initial decay, 
there is {\it no} single $T_2$ for a train of pulses that is
independent of $\tau$.
Our figures show considerable coherence enhancement with just two and
four applied pulses.  Further enhancement can be achieved by
applying more pulses, but this has the cost of requiring more frequent
pulses that must be applied precisely, which has physical limitations.
Isotopic purification can further extend coherence in Si
by eliminating $^{29}$Si nuclei ($^{28}$Si and $^{30}$Si nuclei have no free
spin moments); with $f$ as
the fraction of $^{29}$Si in the bath, we find that
$T_2 \propto 1/\sqrt{f}$ 
(or $T_2 \propto f^{-1/3}$ when pair
contributions dominate the decay of the CPMG sequence).
In principle, it is possible to extend quantum memory time in Si
indefinitely by eliminating all $^{29}$Si,
but this is not possible in GaAs where all isotopes have nonzero spin.

When quantum memory is stored on a nucleus that is identical to the
nuclei in the surrounding bath (rather than a donor), relaxation is
{\it not} suppressed by applying a strong magnetic field and spectral
diffusion is somewhat ambiguous because the concept of the central
spin is ill-defined.  In case (2), we have chosen to study relaxation
of an intrinsic nucleus in a bath of polarized nuclear spins in which
spectral diffusion is suppressed.
We find that this relaxation decay takes
the form of $\exp{\left[-(t / T_1)^2 \right]}$; we find
relaxation times in GaAs of $T_1 = 0.6, 0.7,~\mbox{and}~1.1~\mbox{ms}$ for a central spin of $^{71}$Ga,
$^{69}$Ga, or $^{75}$As, respectively, and relaxation times
in Si of $T_1 = 1.7~\mbox{ms} / \sqrt{f}$ with $f$ as the
fraction of $^{29}$Si ($T_1 \sim 8~\mbox{ms}$ for natural Si in
particular).
These results are relevant, for example, to the
proposal\cite{Taylor03} of 
quantum memory storage in nuclear spin ensembles within quantum dots.
We do not consider loss of memory due to
moving an electron on and off the quantum dot.
We also do not treat the precise dot-dependent decoherence in such 
ensembles that results from spins ``leaking'' off the quantum
dot; such a consideration is made in
Ref.~\onlinecite{Deng05}.  The virtue of our analysis is in its
simplicity and generality.  By considering the rate at which a spin
moves away from a central nucleus, we provide a lower bound for the
coherence times in any such nuclear spin memory proposal in the ideal case
where dipolar interactions among the nuclei dominate decoherence and 
the bath is polarized (otherwise, spectral diffusion also plays a
role).
We study this well-defined problem using 
exact quantum mechanical calculations (i.e., diagonalizing the 
Hamiltonian).



This work was supported by ARO-DTO, ARO-LPS, and NSA-LPS.


\begin{thebibliography}{99}
\bibitem{footnoteCarrPurcell}
The Carr Purcell pulse sequence is more simple
than CPMG but they only differ in the initialization pulse
that is irrelevant for our purpose of storing any initial state.  
The CPMG sequence was designed to provide better protection 
against pulse errors on even echoes.  Since we treat only even multiple-pulse
echoes, having improved coherence over odd echoes in the presence of the
nuclear bath, we use CPMG to denote the sequence.
\end{thebibliography}
\end{document}